\documentclass[a4paper,fleqn]{cas-dc}

\usepackage[authoryear]{natbib}

\begin{document}
\let\WriteBookmarks\relax
\def\floatpagepagefraction{1}
\def\textpagefraction{.001}
\def\z#1{#1}
\def\n#1{\textit{#1}}
\shorttitle{Evolution of Solar-Stellar Activity}
\shortauthors{Maria M. Katsova}

\title [mode = title]{The Evolution of the Solar-Stellar Activity}

\author[1]{Maria M. Katsova}[type=editor,
                        auid=000,bioid=1,
                        prefix=Dr.,
                        role=Researcher,
                        orcid=0000-0001-7511-2910]
\ead{maria@sai.msu.ru}
\ead[url]{www.sai.msu.ru}

\address[1]{Sternberg State Astronomical Institute, 
Lomonosov Moscow State University, Universitetskii prosp. 13, 
119991 Moscow, Russia}

\cortext[cor1]{Corresponding author}

\begin{abstract}
We present \z{a} brief review of observational results \z{contributing} 
to modern ideas on the evolution of stellar activity. 
Basic laws, derived for \z{both} rotation-age and activity-rotation relationships, 
allow us to trace how \z{the} activity of low-mass stars changes with age 
during their stay on the main sequence. 
We focus on \z{the} evaluation of the activity properties of stars 
that \z{could} be analogs of the young Sun. 
Our study includes joint consideration of different tracers of activity, 
rotation and magnetic fields of Sun-like stars of various ages. 
We \z{identify} rotation periods, when the saturated regime of activity 
changes to the unsaturated mode, \z{when} the solar-type activity \z{is formed}: 
for G- and K-\z{type} stars, they are 1.1 and 3.3~days, respectively. 
This corresponds to \z{an} age interval of about $0.2 - 0.6\:$Gyr, when 
\z{regular} sunspot cycle began to be established on the early Sun. 
We discuss properties of the coronal and chromospheric activity in young Suns. 
Our evaluation of the EUV-fluxes in the spectral range of $1350-1750\:$\AA\ 
shows that the far-UV radiation of the early Sun was a factor of 7~\z{times} 
more intense than that of the present-day Sun, and twice higher when the 
regular sunspot cycle was established. 
For the young Sun, we can estimate the possible mass loss rate associated 
with quasi-steady outflow as $10^{-12}\:M_\odot$/yr. 
\z{The} results of \z{observations} of the largest flares on solar-type 
stars \z{are also discussed, leading to} conclusion that the most powerful 
phenomena occur on the fast-rotating stars in the saturated activity regime. 
Our estimate of the stellar magnetic fields makes it possible to evaluate 
the maximal possible flare energy. 
This \z{could help us better} understand the origin of extreme events on 
the Sun in the past.
\end{abstract}


\maketitle

\section{Introduction}

A wide variety of magnetic activity phenomena in different spectral ranges have been observed in stars 
with masses $< 1.5\:M_\odot$ and effective temperatures $\le 6500\:$K \z{over the} past almost 60 years.
Magnetic activity can be traced in all the layers of stellar atmosphere and on various timescales 
(e.g. \citealt{Wilson1978,Baliunas1995,Gudel1997,Lockwood2007}). 
These stars possess a radiative core and an outer convection zone. 
The magnetic field is \z{generated} there \z{via} a dynamo mechanism \z{whose} 
properties depend on the stellar interior structure. 
The interaction \z{between} convection and axial rotation creates magnetic fields, and thereby 
\z{the whole set of} magnetic activity \z{phenomena}, best studied on the Sun due to its proximity. 

Observations of different tracers of stellar activity show\-ed that the main factor 
determining the activity level is the axial rotation of a star. 
\citet{Skumanich1972} was the first who discovered a specific rotational braking law for 
solar-type stars in the Pleiades, Ursa Major, and Hyades open clusters and the Sun from an 
analysis of the time-scales for Ca~II emission decay, rotational braking, and lithium depletion. 
He determined that the rotational decay of all these parameters follows a square-root relation, 
and found that the projected rotational velocities $v \sin i$ of G-type stars on the 
main sequence decrease with age $t$ as $v \sin i \propto t^{-1/2}$.
This rotational braking is caused by the torque provided by the magnetized stellar wind, 
outflowing along magnetic field lines, which is able to efficiently remove the 
angular momentum of the star. 

This relation, called the "Skumanich law", has served as the basis of the \n{gyrochronology method} 
\citep{Barnes2003}, which yields age estimates based on rotation observations.
As \z{it is} known, an intensity of the Ca~II K~line on the Sun varies linearly 
with surface magnetic field strength \citep{Frazier1970}, therefore the stellar Ca~II emission 
can be identified with the (average) surface magnetic field. 
\z{Later on \protect\citet{Schrijver1989} 
found a ratio between the chromospheric Ca~II emission and the absolute value 
of the mean magnetic flux densities: the Ca~II HK flux is proportional to 
$(f{\!B})^{0.5}$ (where $f$ is an area filling factor and $B$ is the intrinsic 
field strength.} Then this law (Fig.1 in \citealt{Skumanich1972}) 
implies that the average surface (dynamo) field is 
proportional to the rotational velocity and decays as the inverses square 
root of the time \z{while} the star is located on the main sequence. 

Now this conclusion is confirmed by direct measurements of magnetic fields 
in low-mass stars $(\sim 0.1 - 2 M_\odot)$ of different ages 
with spectropolarimetric methods \citep{Vidotto2014}. 
They empirically found that the unsigned average large-scale surface field 
$\langle|B_V|\rangle$ 
is related to age as 
$t^{0.655\pm 0.045}$. 
This relation\z{ship} between the magnetic field and age could be used as 
a way to estimate stellar ages ("magnetochronology"), although it would not 
provide a better precision than most of currently adopted age-dating methods. 
Results by \citet{Rosen2016} showed a significant decrease in the magnetic field strength 
and energy as the stellar age increases from 100 Myr to 250 Myr, 
while there is no significant 
age dependence of the mean magnetic field strength for stars with ages 
$250-650\:$Myr. 
The spread in the mean field strength between different stars is comparable to 
the scatter between different observations of individual stars. 
They applied a modern Zeeman-Doppler imaging (ZDI) code in order to reconstruct 
the magnetic topology of all stars (see also \citet{Kochukhov2017}).

Note that the ZDI measured field is a small fraction of the total 
magnetic field, and a fraction which varies with rotation. Now 
\citet{Kochukhov2020} developed a new magnetic field diagnostic 
method based on relative Zeeman intensification of optical atomic 
lines with different 
magnetic sensitivity. They found that the average magnetic field 
strength $Bf$ drops from $1.3-2.0\:$kG in stars younger than about 
120~Myr to $0.2-0.8\:$kG in older stars. These results suggest 
that magnetic regions have roughly the same local field strength 
$B\approx 3.2\:$kG in all stars, with 
the filling factor $f$ of these regions systematically increasing with 
stellar activity. Comparison of these results with the 
spectropolarimetric analyses of global magnetic fields in the same young 
solar analogues shows that ZDI recovers about 1\%\ of the total magnetic 
field energy in the most active stars. These new data are very important 
for understanding of different evolutionary phases of the solar-like 
magnetic dynamo and can be used to estimate magnetic characteristics of 
the Sun during the first $\sim 1\:$Gyr of its life.

The rotation periods of young stars in star formation regions in the 
stage of gravitational contraction were determined directly from 
observations of the rotational modulation of their photometry. The 
rotation periods of young stars with masses from 0.8 to $1.2\:M_\odot$ 
were found to vary from 7 days to about $< 1\:$day as the age varied 
from 1 to 70~Myr (Fig. 7 in \citealt{Messina2011}), i.e. rotation 
accelerated as the star approached the main sequence. The subsequent 
braking of rotation associated with the angular momentum loss due to the 
magnetized stellar wind occurs over significantly larger time scales of 
billions of years.

Gyrochronology is based on the observation that main-sequence stars spin 
down as they age \citep{Barnes2003}, and thus if correctly calibrated, 
rotation can act as a reliable determinant of their ages. To calibrate 
gyrochronology, the relationship between rotation period and age must be 
determined for cool stars of different masses, which is best 
accomplished with rotation period measurements for stars in open 
clusters with well-known ages \citep{Meibom2015}. Then gyrochronology 
may be a more precise clock than other techniques like asteroseismology 
and isochronal methods for cool main-sequence stars.

Observations of rotation of open cluster members of different ages 
revealed two populations: \z{the} fast rotating stars and the slower 
rotators. One of conclusions by Barnes' PhD thesis \citep{Barnes1998} was: 
the existence of the ultra-fast rotators implies that angular momentum 
loss from young stars is inconsistent with a Skumanich-type slowdown. It 
may be interpreted either as evidence for magnetic saturation at high 
rotation rates or for a different magnetic field configuration in young 
stars, before the Skumanich spindown phase. Moreover, with increasing 
age, the number of rapidly rotating stars decreased \citep{Barnes2003}. 
Such coexistence both fast and slow rotators among young low-mass stars 
was discovered later on during the Kepler mission data for more than 
34~000~stars \citep{McQuillan2014}. 

Later, \citet{Mamajek2008} proposed a way for improved age estimation 
for solar-type dwarfs using acti\-vi\-ty-rotation diagnostics. Their new 
activity-age calibration has typical age precision of $\sim 0.2\:$dex
for these normal F7--K2 dwarfs aged between the Hyades and the Sun 
($\sim 0.6-4.5\:$Gyr). They showed that the coronal activity index as 
measured through the $R_X = L_X/L_{bol}$ (the X-ray to the bolometric 
luminosity ratio) has nearly the same age- and rotation-inferring 
capability as the analogous chromospheric activity index measured 
through the $R^{'}_\textit{HK} = L_\textit{HK}/L_{bol}$ 
(the chromospheric to the total bolometric \z{luminosity ratio}). 
It is calculated from the so-called 
S-index, a band ratio measurement of the Ca~II H and K~emission 
line strength at 3933\,\AA\ and 3968\,\AA, \z{respectively}, from which 
the underlying stellar photospheric contribution is then subtracted. 
Physically, variability in these lines is a time manifestation of 
stellar surface magnetic inhomogeneities, and is often used as a proxy 
for the stellar magnetic activity \citep{Baliunas1996}. One use has been 
in long-term studies of chromospheric activity, aimed at finding 
starspot cycles. This monitoring continues for nearly 60~years for 
\z{111} stars starting with the HK-Project at the Mt Wilson observatory 
\citep{Wilson1978,Noyes1984,Baliunas1995,Henry1996,Wright2004,Arriagada2011}.

Thus, \z{by} systematic observations of different activity indices for 
stars of various ages, \z{the} gyrochronology method (rotation-age dependence) 
was developed, \z{making} it possible to explore the evolution of 
activity of a star over its life on the main sequence.

\section{Results}

\subsection{A scenario of the evolution of stellar activity}

Available observations permit us to compare the main activity indices of 
large numbers of stars of various ages (rotation rates) \z{and} to 
understand how activity evolves. \z{A major expansion in the exploration 
of} the evolution of stellar activity began through use of data from 
\n{the California, Carnegie and Magellan planet search programs} 
\citep{Wright2004,Arriagada2011}. They provided data on the 
chromospheric activity (index $R^{'}_\textit{HK}$) for more than 1300 
northern and southern stars obtained as a by-product of these projects. 

The X-ray data for coronae of these stars (the coronal activity index 
$R_X$) were derived from observations in the soft X-rays carried out 
with the \n{ROSAT} and \n{XMM-Newton} (for instance, 
\citealt{Hunsch1999,Schmitt2004}), and for a few dozen stars from results by 
\citet{Pop2010,Pop2011}. This dataset can be used to study the location 
of solar activity among activity \z{phenomena} occurring on other late-type 
stars, and to trace the evolution of solar-like activity from ages 
of about $100\:$Myr to $> 6\:$Gyr. 

\begin{figure*}
	\centering
		\includegraphics[width=.85\textwidth]{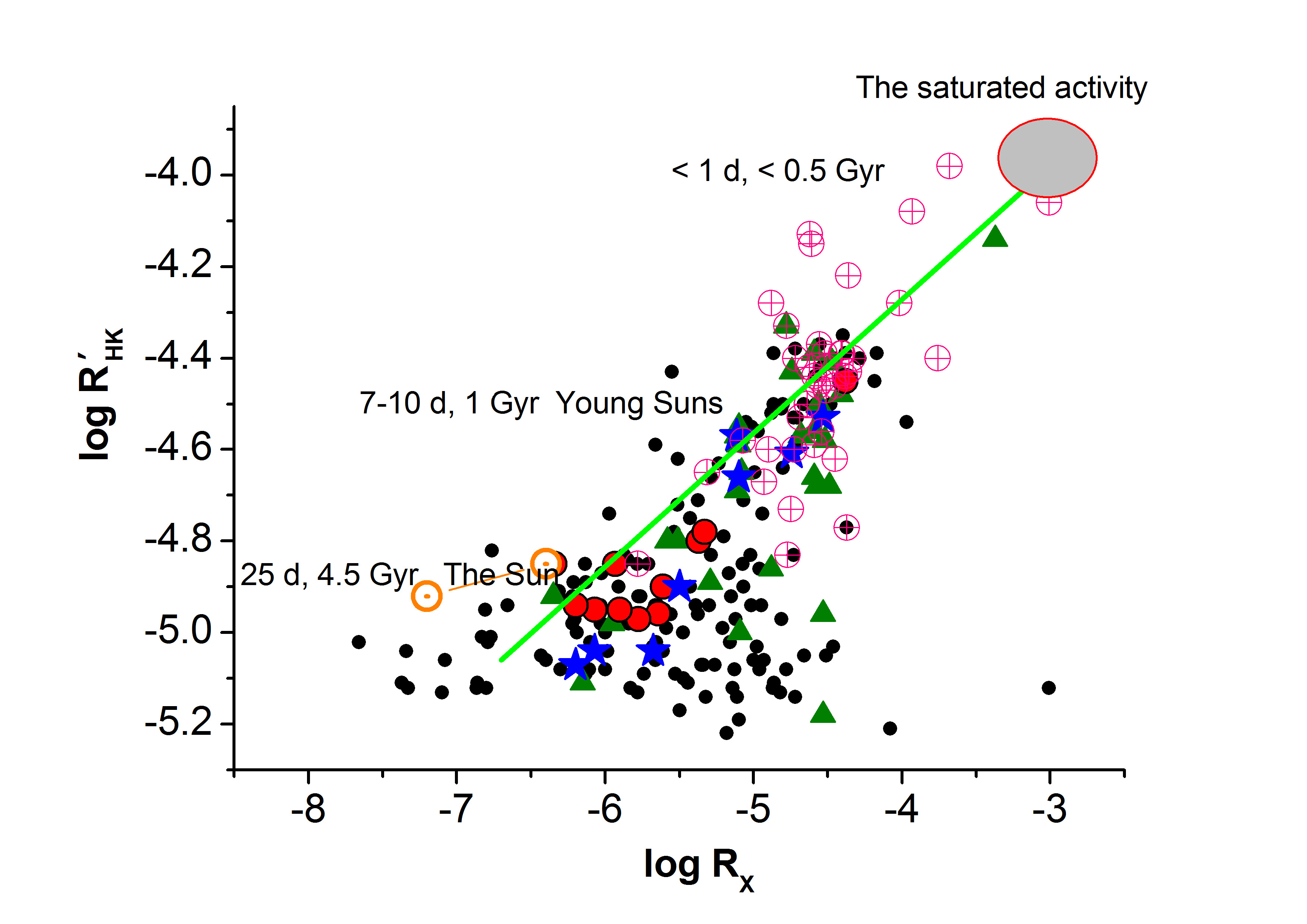}
	\caption{The "chromosphere -- corona" diagram for F-, G- and K-type 
    main-sequence stars. ~
    \n{Notes:} ~ the grey ellipse marks \z{conditionally the area where 
    the stars with saturated activity regime are located;} green 
    triangles (28 stars) and magenta circled crosses (48 stars) are 
    stars with detected lithium abundances 
    \protect\citep{Montes2001,Lopez2010,Maldonado2010,Mishenina2012}.
    Red circles (\z{11 stars}) and blue asterisks (8 stars) note the 
	HK-Project stars with \n{Excellent} and \n{Good} \z{starspot} cycles 
    \protect\citep{Baliunas1995}; black dots (158 objects) are the stars 
    discovered during Exoplanet Search Programs. The green straight line 
    corresponds to the ratio for \z{the} gyrochronology of 
    \protect\citet{Mamajek2008}.
    The Sun at the maximum and minimum of the sunspot cycle 
	is marked by an orange solar symbol.
}
	\label{FIG:1}
\end{figure*}

Considering the indices of coronal and chromospheric activity together, 
we proposed the "chromosphere-corona" diagram: ($R^{'}_\textit{HK}$ 
versus $R_X$) which allowed us to compare 
a level of the solar activity with that 
of other stars, and to analyze how these parameters change 
\citep{KaLi2011,Katsova2012}. 
This diagram \z{is presented} in Fig.\ref{FIG:1}; 
it \z{includes} over 250 stars from several \z{observational programs}. 
Besides stars from the Mt Wilson HK-Project with chromospheric 
\z{starspots} cycles determined as \n{Excellent} and \n{Good} 
\citep{Baliunas1995} and stars from the above mentioned exoplanet search 
program, we added stars with detectable lithium abundances 
\citep{Lopez2010,Maldonado2010,Mishenina2012}. 
\z{The lithium data, including 
our own observations, were added because the Li~I 6708$\:$\AA\  line is 
known as an indicator of stellar age and, consequently, the activity 
level \protect\citep{Skumanich1972}. The lithium abundance is sensitive 
to the temperature near the bottom of the convection zone: lithium is 
efficiently depleted when this temperature reaches about 2-2.5$\:$MK. 
Physical conditions in these deep layers determine the properties of the 
dynamo and, in general, the features of the magnetic field generation. 
We discovered the correlations between the Li abundances, rotational 
velocities and the level of the chromospheric activity for F-, G- and 
K-type main-sequence stars and found that it was tighter for the stars 
slightly cooler than the Sun, with the effective temperatures 
$5700>T_{\textit{eff}}>5200\:$K. 
For highly active stars, we confirmed that both the Li 
abundance and the activity level are determined by the age-dependent 
rotation rate \protect\citep{Mishenina2012}. The stars with detectable 
Li in the "chromosphere-corona" diagram cover a wide range of the 
activity indices up to the saturation level. Apparently, differences in 
behavior of the Li abundances in the stars of various activity levels 
could be associated with the changes of a relative contribution of the 
small-scale (local) and large-scale magnetic fields in generating the 
activity \protect\citep{KaLiMi2013}.} In particular, this is the case 
when the Li excess can be evidence for higher stellar activity 
associated with very powerful stellar flares (see, for example, results 
by \citet{Livshits1997,Living1997,Ramaty2000} for solar flares, and 
those by \citet{Montes1998} for a stellar flare).

\z{The green straight line corresponds to the linear regression between 
the coronal and chromospheric indices for all intervals of their values 
(the ratio for the} one-parameter gyrochronology) by 
\citet{Mamajek2008}: 
\begin{equation}
 log R^{'}_\textit{HK} = - 4.54 + 0.289 (\log R_X + 4.92).
\end{equation}

These authors used the updated ages of well-studied nearby open clusters 
(e.g., alpha Perseus, Pleiades, Hyades) and considered the 
rotation-activity-age relation. Their new activity-age calibration has 
typical age precision of $\sim 0.2\:$dex for normal solar-type dwarfs aged 
between the Hyades and the Sun ($ \sim 0.6 - 4.5\:$Gyr). This \z{allowed}
 us to \z{determine a region} on this diagram where the youngest stars 
with saturated activity are located, the approximate location of fast 
rotators with ages $<0.5\:$Gyr, ($log R^{'}_\textit{HK} \sim -4.0$ and 
rotation periods $P_{rot} < 1\:$d), the level of the young Sun in the 
age of $\le 1\:$Gyr ($\log R^{'}_\textit{HK} \sim -4.5$, $P_{rot}$ around 
$7-10\:$d), and a place of the contemporary Sun in the age 4.5~Gyr 
($\log R'_\textit{HK} \sim -4.88$, $P_{rot}=25\:$d) as well.

A few features in the behavior of different groups of stars \z{should be 
noted. For example,} the chromospheric activity of the Sun is clearly 
higher than that for the vast majority of stars in the solar vicinity 
(e.g. the same ages), while the solar corona is much weaker even at the 
maximum of the sunspot cycle as compared to coronae of other main-sequence 
stars. The stars with well-defined cycles situate along the 
line corresponding to the \n{one-parameter gyrochronology} by 
\citet{Mamajek2008}. Their chromospheric and coronal activity decreases 
almost simultaneously. This kind of activity follows Skumanich' law and 
relates to \n{solar-type activity} that implies a starspot cycle 
formation. 

From the other side, stars with the detectable \z{Li}, which are more 
active and younger than others among these objects, deviate from this 
line: there is also a significant group of stars for which the 
chromospheric activity diminishes, \z{while} their coronal radiation 
spans a wide range or remains quite intensive. \z{Note that a wide 
spread in $R'_\textit{HK}$ for a given $R_X$ for low $R_X$ is partly because 
$R'_\textit{HK}$ is uncalibrated for metallicity (the iron abundance), while 
$R_X$ usually is \citep{Saar2012}.} 

The upper part of this diagram (the grey ellipse) indicates direction to 
a region where the youngest, fast-rotators with very high activity are 
located. These stars are characterized by saturated activity \z{and even 
supersaturated regimes for the fastest rotators.} This diagram shows 
possible paths of the evolution of solar-type activity: decay of the 
chromospheric and coronal activity can occur by different ways. 
Moreover, in fact, it represents the temporal behaviour of solar-type 
activity, namely, how the starspot cycle becomes more pronounced as the 
star decelerates and the chromosphere weakens. 

When the rotation of a young star slows, the chromospheric and the 
coronal activity seems to decline synchro\-nously. The solar-like activity 
of the most F- and early G-\z{type} main-sequence stars evolve by this 
path. However, the activity of lower mass stars, from G5 to K7, after a 
certain point evolves \z{differently}: the chromospheric activity 
diminishes to the solar level, while coronae stay stronger than that of 
the Sun. Two possible paths of the evolution of activity can be 
associated with the different depth of the convective zone of these 
stars \citep{KaLiMi2013}. Physically, this means that the relative 
contribution of local (small-scale) and large-scale magnetic fields to 
activity differs for F-, G- and K-\z{type} stars. The solar-like dynamo 
tells us that the large-scale magnetic fields are generated in the bottom 
of the convection zone, in the tachocline, while the small-scale fields 
are originated in subphotospheric layers. Therefore the location of the 
lower boundary of the convection zone is important parameter.

\subsection{On the activity-rotation relationship and different 
regimes of stellar activity}

The first evidence for the dynamo-induced nature of stellar coronal 
activity was obtained by \citet{Pallavicini1981}, who found that the 
X-ray luminosity \z{in} accord\z{ance} with Einstein observations scaled 
with \z{the} projected rotational velocity as 
$L_X \propto (v \sin i)^{1.9\pm 0.5}$ (or $L_X \propto P_{rot}^{-2}$, 
where $P_{rot}$ is the rotation period of a star), and no dependence on 
bolometric luminosity. This relationship between rotation and activity 
was found to break down when X-ray luminosity reaches a saturation level 
of $L_X \sim 10^{-3} L_{bol}$ \citep{Micela1985} independent on spectral 
type (increasing with decreasing bolometric luminosity). This saturation 
level is reached at a rotation period that increases toward later 
spectral types \citep{Pizzolato2003}. It was unclear what a cause of 
this saturation is: a saturation of the dynamo itself, or a saturation 
of the filling factor of active regions on the stellar surface 
\citep{Vilhu1984}. But once saturation occurs, the X-ray emission becomes a 
function of only the bolometric luminosity \citep{Pizzolato2003}, or 
effectively the mass, color, or radius of the main-sequence star. 

Rather than the rotation period, studies of the rotation-activity 
relationship frequently use the Rossby number $Ro=P_{rot}/\tau_{conv}$, 
which is the ratio of the stellar rotation period $P_{rot}$ to the 
convective turnover time $\tau_{conv}$  in that part of the convection 
zone where dynamo activity is situated. This dimensionless \z{value} is 
an important parameter of astrophysical dynamo for characteristics of 
convective motions in a stellar convection zone. It depends on both 
axial rotation rate and spectral type, and is a measure of the 
importance of Coriolis forces in introducing helicity into convective 
motions \citep{Noyes1984}.

The largest sample to date (more than 800 solar- and late-type stars, 
including both field stars and stars in nearby open clusters of ages 
$40-700\:$Myr) with well-measured X-ray luminosities and photometric 
rotation periods has been adopted from the literature by 
\citet{Wright2011}. An analysis of this consolidated catalogue showed 
that the relation between rotation (in the form of the Rossby number, 
$Ro=P_{rot}/\tau$) and stellar activity (in the form of the X-ray to 
bolometric luminosity ratio, $R_X=L_X/L_{bol}$) can be divided into the 
unsaturated, saturated, and even supersaturated regimes for the fastest 
rotators of the coronal X-ray emission \citep{Wright2011,Reiners2014}. 

Stellar activity can be in \z{the saturated regime} with the coronal 
index $\log R_X \approx -3$, \z{a} state of young stars in which 
activity is independent of rotation. High-level irregular, chaotic 
activity of these stars evolves eventually into another regime, a second 
mode, \z{solar-type activity}, which strongly depends on the rotation 
period. In this regime, coronal emission can be as low as the current 
quiet Sun ($\log R_X=-7.1$ in the minimum of the solar cycle; e.g., 
\citealt{Peres2000}). As the name implies, in particular, formation of a 
more or less regular cycle is possible \z{for} such typical active 
phenomena as starspots, active regions, flares and \z{coronal mass 
ejections} (CMEs). 

\z{Although there are some stars in the saturated regime -- e.g., AB~Dor,
LQ~Hya, with photometric variability where multi-periodicity were 
revealed. Perhaps, saturated stars just have spot cycles, but not plage 
cycles, they surfaces too full with active regions, and hence there is 
no activity cycle in either the chromosphere or the corona. Note that we 
are talking about regular steady cycles in all tracers of activity. But 
sometimes it is difficult to separate "true" cycles from stochastic 
variations, because the resolved time-scales of stellar activity are 
insufficient to decide reliably that a cyclic variation for a particular 
star is similar to the well-known 11-yr sunspot cycles. We carried out 
the wavelet analysis of the longest available stellar activity record -- 
photometric monitoring of a young star V833~Tau for the 120~yr; it 
reveals that variations obeys the continuous spectrum of fluctuations 
without any significantly pronounced peaks. We find that the observed 
variations of V833~Tau with time-scales of $2-50\:$yr should be 
comparable with the known quasi-periodic solar mid-term variations, 
whereas the true cycle of V833~Tau, if it exists, should be of about a 
century or even longer \protect\citep{Stepanov2020}.}

\begin{figure*}
	\centering
	\includegraphics[width=.75\textwidth]{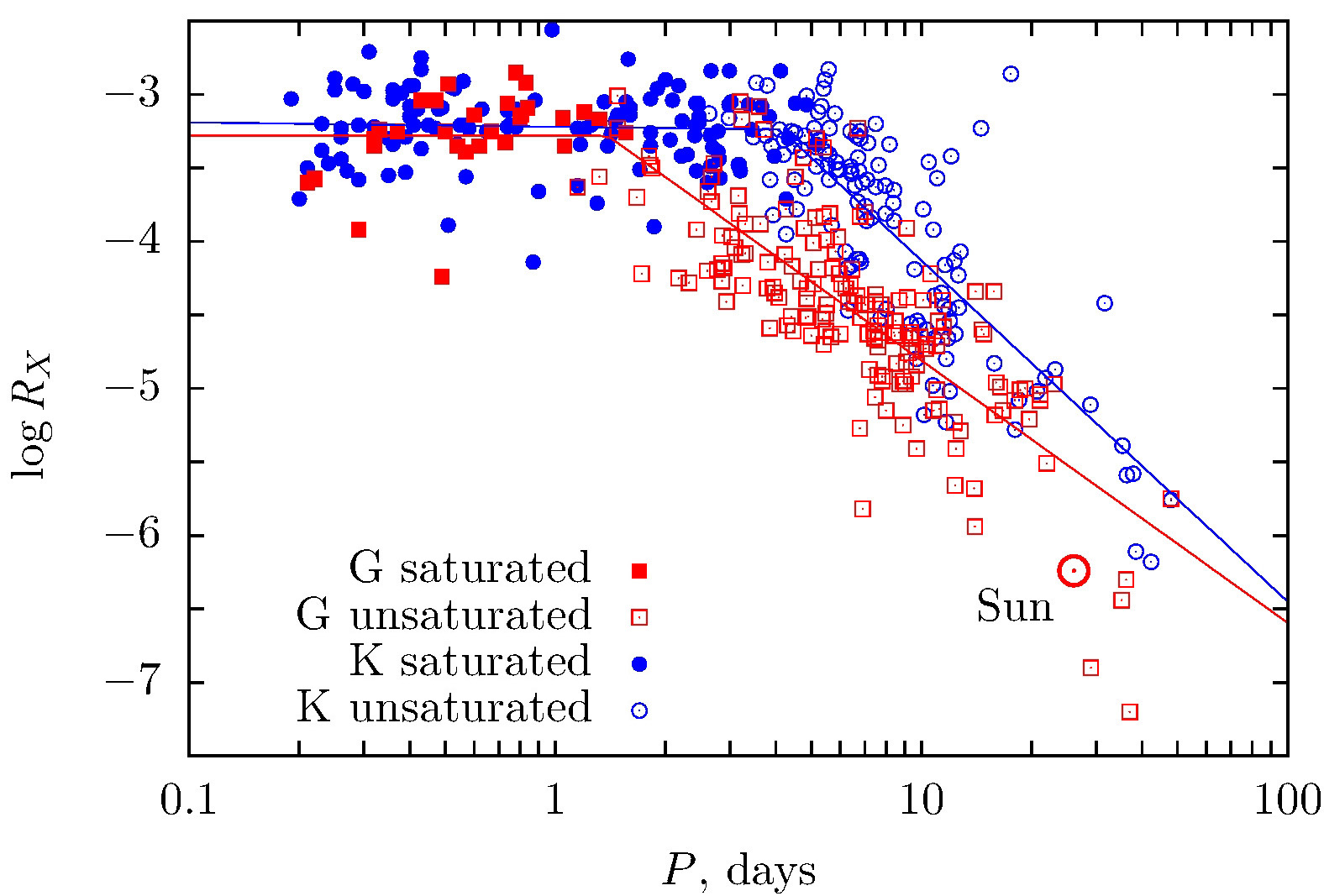}
	\caption{X-ray to bolometric luminosity ratio versus rotation 
    period, for G- and K-\z{type} stars 
	extracted from the catalog by \protect\citet{Wright2011}.
	G- and K-\z{type} stars are marked by red and blue \z{symbols},
	respectively: filled \z{symbols} relate to 
	the saturated activity regime, empty \z{symbols} 
	denote the unsaturated mode of activity. 
	The location of the Sun is marked by its \z{symbol}.}
	\label{FIG:2}
\end{figure*}

The evolution of activity in both regimes are traced separately for 
G- and K-type stars in Fig.\ref{FIG:2}, where the coronal indices 
are plotted against rotation periods, based on data extracted from 
the catalog by \citet{Wright2011}.

As seen in Fig.\ref{FIG:2}, the period span of the saturated 
activity epoch for G-type stars is shorter than that of K-type
ones, and the transition from saturation to unsaturated regime occurs at 
different rotation periods for G- and K-type stars. In order to 
clarify in more detail when the saturated regime changes to the 
unsaturated mode of solar-type activity, we carried out more refined 
analysis \citep{Nizamov2017} of the same dataset as in \citet{Reiners2014}. 
We considered G-, K- and M-type dwarfs separately and showed 
that the transition from the saturated activity mode to the solar-like 
one takes place at the rotation periods of 1.1, 3.3 and 7.2~days 
for G2-, K4- and M3-type stars, respectively. 

This result allows us to estimate the time interval, when the sunspot 
cycle could appear in the Sun. It appears, that for solar-type 
stars, the saturated activity epoch ends at the ages of $0.2-0.6\:$Gyr. 
After this age, the solar-type activity regime begins to be established, 
and conditions for formation of a starspot cycle are created. Thus, the 
sunspot cycle could initiate itself on the Sun at periods of 
$1-3\:$days, and that the epoch of the solar-type activity covers a wide 
interval of the rotation periods and can last from the ages of hundred 
millions to a few billions of years.

Recently \protect\citet{Curtis2019} found that stars cooler than G2 
start to slow their spindown, with the slowing increasing with lower 
mass, i.e. K-type dwarfs appear to spin down more slowly than F- and 
G-type dwarfs. This means that gyrochronology only works for solar-type 
($\sim$G2) stars and more massive stars. These changes in activity regimes 
may be due to different dynamo configurations.

\subsection{Properties of Activity of the Young Sun}

According to modern ideas on the internal structure and evolution of 
solar-mass stars, there are differences between early stage of evolution 
and epoch when a star is located on the main sequence. So, parameters of 
a star on a way toward the main sequence in pre-main sequence (pms) 
stages change much faster than those during its further life on the main 
sequence. During the \z{first} pms-stage lasting a few millions of 
years, a fully convective star turns into a normal, almost \z{steady} 
star with the radiative core and convective envelope. Then after 
\z{about} 50~Myr the main-sequence stage arrives, which continues for 
about 9~Gyr \citep{Baturin2017}. Now our Sun is a main-sequence G2~V 
star of intermediate age, 4.6~Gyr old, with the effective temperature of 
$\sim 5775\:$K. According to the standard model 
(e.g., \citealt{Bahcall2001}), 
the current solar luminosity has increased by 30\%\  as compared to that 
on the pms, and its radius and the depth of  the convection zone are 
10\%\  larger. 

The lifetime of the Sun on the main sequence can be divided \z{in} 
a few epochs: the early Sun when the Solar Planetary System just was
forming, when the solar rotation rate was $10-20$ times faster than
nowadays, the era of the young Sun when a cycle was established 
(the rotation rate was $2-5$ times faster), and the contemporary epoch
of the slow-rotating Sun.

Activity in solar-like stars over this wide range of ages has been
investigated in \z{\n{"The Sun-in-Time"}} program, started almost 
30~years ago \citep{Gudel1997,Ribas2005}. This multi-wavelength project 
aimed to trace changes in the coronal and chromospheric emission over 
the lifetime of the Sun on the main sequence. It was useful, in 
particular, for studying effects of high energy radiation on exoplanets 
around the stars. 

Beside the results of the long-term monitoring of the stellar 
chromospheres begun by the Mt Wilson HK-Project mentioned above, the 
main data sources for activity diagnostics in the chromospheres and 
transition regions were several space missions, in particular, 
\n{International Ultraviolet Exporer (IUE)} satellite, \n{Extreme 
Ultravioler Explorer (EUVE)} and \n{Hubble Space Telescope (HST)}. 
Extensive research of stellar coronae was carried out using the 
\n{ROSAT, Einstein, ASCA, Chandra} and \n{XMM-Newton} 
(see \citet{Guinan2009} for details). Five solar analogs 
(G0~V-G5~V stars) of different ages were selected for their program, 
and the collected data allowed them to estimate the physical parameters
of activity of the young Suns. The \z{following} stars were selected: 
EK~Dra (G1.5~V, with age of 130~Myr) with the rotation period 
$P_{rot}= 2.8\:$d, 
$\pi^1$~UMa (G0.5~V, 300~Myr) with $P_{rot} = 5\:$d, 
$\kappa^1$~Cet (G5~V, 750~Myr) with $P_{rot}=9.3\:$d, 
$\beta$~Com (G0~V, 1.6~Gyr) with $P_{rot}=12\:$d, and 
$\beta$~Hyi (G0~V, 6.7~Gyr) \citep{Gudel1997,Ribas2005}. 

Results of the "The Sun in Time" program allowed us to \z{assess the} 
physical conditions in coronae of the youngest G-\z{type} stars, which 
are very powerful: the X-ray luminosity, $L_X$, is up to a few times of 
$10^{30}\:$erg/s, 
coronal temperatures reaches $10^7\:$K, electron densities at 
the base of the corona are $3-5\times 10^9\:$cm$^{-3}$, and the emission 
measure is $10-30$ times higher than that in active regions on the 
contemporary Sun. For example, the coronal indices for EK~Dra and 
$\kappa^1$~Cet are $\log R_X=-3$ and $-4.4$ respectively, indicating 
that the X-ray luminosity of the young Sun exceeded that from the 
present-day active Sun by $3-4$ orders of the magnitude. 
In addition, with age, the coronal emission became cooler 
\citep{Ribas2010}.

\z{In order to understand} the impact of the young Sun on the origin of 
the biosphere and the geological history of the Earth, it is important 
\z{to evaluate} the intensity of the EUV radiation, quasi-steady 
outflows of the plasma and fluxes of high energy accelerated particles 
affecting the radiation environment in that era. The stellar UV excess 
is roughly proportional to the chromospheric index $R'_\textit{HK}$ or 
\z{to} stellar age, at least for stars younger than 1~Gyr \citep{Findeisen2011}. 
We \z{have} selected here 15 fast-rotating solar-like G-type stars 
of approximately \z{that} age for estimation their 
EUV-fluxes. \z{We have used} data obtained with \n{GALEX (Galaxy Evolution 
Explorer)} in two wide spectral ranges: Near-UV $1750-2750\:$\AA\  and 
Far-UV $1350-1750\:$\AA\  \citep{Bianchi2011}. The FUV-range is more 
sensitive to activity level and is spared from \z{the} problems with the 
flux saturation. For these 15~stars, we collected the \n{GALEX} 
FUV-fluxes from \citet{Bianchi2011}, and then recomputed these fluxes at 
the distance 1 a.u. in erg/(cm$^2\,$s), then calculated the star-to-Sun 
flux ratio (Table \ref{tbl1}).

\def\s{\hphantom{0}}

\begin{table*}[width=.9\textwidth,cols=7,pos=h]
\caption{
The \n{GALEX} FUV-fluxes for 15 solar-like G-type stars.
The table contains the names of the stars (1), spectral types (2) 
and parallaxes, in milliarcseconds (3), the measured \n{GALEX} 
FUV-fluxes, in microJansky (4) 
from \protect\citet{Bianchi2011}, FUV-fluxes, recalculated 
to the distance 1 a.u., 
in erg/(cm$^2$ s), (5), and the star-to the Sun flux ratio (6).
}\label{tbl1}
\begin{tabular*}{\tblwidth}{@{} RLLC@{}CCC@{} }
\toprule
N & Star & Spectral & Parallax, & Flux, & Flux at 1au, & Flux ratio \\
  & Name & type	    & milliarcseconds	& microJansky	& erg/(cm$^2$ s) 
  & Star/Sun \\
\midrule
 & 1 & 2 & 3 & 4 & 5 & 6 \\
\midrule

\rule{0pt}{17pt}
1  &  HD 224540  &  G0     &  22.40  &  15.77  &  \s{}6.79   &  3.40 \\
2  &  HD 138159  &  G3 V   &  14.83  &  16.67  &  16.39  &  8.20 \\
3  &  HD 19423   &  G2 V   &  19.51  &  20.81  &  11.82  &  5.91 \\
4  &  HD 212619  &  G3 V   &  14.86  &  17.08  &  16.72  &  8.36 \\
5  &  HD 131179  &  G5/6 V &  25.26  &  33.09  &  11.21  &  5.61 \\[6pt]

6  &  HD 218614  &  G5 V   &  18.66  &  27.50  &  17.06  &  8.53 \\
7  &  HD 221343  &  G2 V   &  19.76  &  29.22  &  16.18  &  8.09 \\
8  &  HD 119824  &  G0     &  21.34  &  20.69  &  \s{}9.82   &  4.91 \\
9  &  HD 12264   &  G5 V   &  24.46  &  44.65  &  16.13  &  8.07 \\
10 &  HD 214867  &  G3 V   &  16.77  &  17.46  &  13.42  &  6.71 \\[6pt]

11 &  HD 109360  &  G5     &  15.33  &  10.21  &  \s{}9.38  &  4.69 \\
12 &  HD 211847  &  G5 V   &  20.49  &  31.78  &  16.36  &  8.18 \\
13 &  CD-33 15016&  G2     &  \s{}9.80   &  64.86  &  14.61  &  7.31 \\
14 &  HD 222628  &  G2 V   &  10.60  &  63.90  &  12.31  &  6.16 \\
15 &  HD 30386   &  G3 V   &  10.94  &  10.79  &  19.48  &  9.74 \\

\bottomrule
\end{tabular*}
\end{table*}

We obtained that the mean FUV-flux is $13.9 \pm 3.5 \:$ erg/(cm$^2\,$s)
at the distance of 1 a.u. for 15 young main-sequence G-type
stars; this value exceeds the FUV-flux of the contemporary Sun by the 
factor of almost 7. For a proxy of the Young Sun, comparison of 
$\kappa^1\:$Cet spectra, given by \citet{Ribas2010}, with the solar one 
showed that their contrast in the UV continuum ($1000-1700\:$\AA) is 
equal to 2. This means that the young Sun irradiated the terrestrial 
surface \z{in the UV range several times more intense} in the era of 
formation of the biosphere, when life arose on the Earth, than today.

\subsection{On magnetic fields, mass losses, and CMEs of Young Suns}

Magnetic field generation in cool stars is caused by a dynamo process 
which is primarily an interaction between rotation (and its shear) and 
convection. Magnetic activity becomes stronger \z{with} faster stellar 
rotation. As mentioned above, magnetic fields decrease with stellar age, 
as we pass from fast rotators to slowly rotating stars.
 
In order to find out the energetics of quasi- and non-steady processes 
on the young Sun, which occurred when the regular sunspot cycle was 
established, we need magnetic field observations. These measurements 
were carried out in the frameworks of the spectropolarimetric \n{Bcool}
project magnetic survey of 170 solar-type stars by \citet{Marsden2014}. 
The detected surface magnetic fields were found on 67 objects. \z{It} 
was \z{established}, that in general, the mean value of the strength of 
the magnetic field $|B_l|_{mean}$ increases with rotation velocity and 
\z{decreases} with age. For all G-\z{type} dwarf star samples, the mean 
value of the strength of the magnetic field $|B_l|$ was found 3.2~Gauss. 
In particular, the modulus of the longitudinal magnetic field for young 
solar analog $\kappa^1\:$Cet is equal to 7.7~Gauss. \z{It is important 
to note that these ZDI field measurements detect only the residual 
large-scale fields after dominant small-scale cancellation effects 
\protect\citep{Kochukhov2020}.} 

For evaluation of the magnetic field of the Young Sun, we selected
samples of G-\z{type} dwarf stars with values $|B_l|$, exceeding 
$3\sigma$, excluding the youngest, fast rotating stars. For the dozens 
of G-\z{type} stars with the rotation period $P_{rot}=7\:$days, we 
obtained this averaged value as $|B_l|=4.72\pm 0.53\:$Gauss. We compared 
this value with the global magnetic field of the contemporary Sun as a 
star near the maximum of the sunspot cycle. The averaged over the 
Carrington rotation the magnetic fields of the Sun as a stars at high 
activity level (for example, in 1991), is $|B_l|=0.5\:$Gauss 
\citep{Kotov1999}. From here we can conclude that the average magnetic 
field strength of the young Sun, when the sunspot regular cycle arose, 
was at least an order of magnitude higher than that for the maximal Sun 
in the modern era \citep{KaLi2014}. Note also that the local magnetic 
flux in plage of young low-mass stars can reach $3-5\:$kG, and active 
regions can cover up to $20-30\:$\% of the stellar surface 
\citep{Saar2001,Reiners2008}.

The rotational evolution of low-mass stars is governed by an angular 
momentum loss caused by the interaction of the stars's magnetic field 
with its ionized wind. \z{It is potentially possible to} evaluate the 
mass loss of a star of a given age, in the case when its rotation rate 
is known. This \z{opens} a way of find\z{ing} out the mass loss \z{rate} 
by the young Sun. \z{This only works if the angular momentum loss theory 
is correct.}

The plasma in the wind escapes its host star at the distance where 
the wind reaches the Alfven speed $v_A=B/(4\pi \rho)^{1/2}$, where $B$ 
is the local magnetic field strength, and $\rho$ is the local mass 
density. $\rho = n m_p$ is the mass density ($n$ is the number density 
and $m_p$ is the proton mass). Beyond this surface, the wind velocity 
exceeds the Alfven speed, and the wind is no longer in contact with the 
star via magnetic fields.

As the first step, the mass loss can be estimated from the expression by 
\citet{Weber1967} relating mass and angular momentum loss. The total 
angular momentum loss rate is $dJ/dt = -2/3 \Omega \dot M R^2_A$, 
where $dJ/dt$ is the angular momentum loss rate, $\Omega$ is the angular 
velocity, $\dot M$ is the mass loss rate, $R_A$ is the average distance 
from the center of the star to a given point on the Alfven surface, and 
where a constant radial magnetic field is assumed at the surface of the 
star. This expression is good for a case of a spherically-symmetric 
wind, a magnetic field that is close to a uniformly distributed one 
across the Alfven surface, and constant moment of inertia for the star 
(i.e., one assumes the time scale for angular momentum loss in this 
expression is short relative to the evolution of the star's internal 
structure). This description breaks down for more complex magnetic 
topologies. Note also that this is appropriate for rotating 
main-sequence stars when their mass and radius do not vary with time. 

The mass loss rate can be also evaluated from the following equation 
for the total torque of the star $\dot M = \tau_w / \Omega R^2_A$,
where $\tau_w$ is the stellar torque, $\Omega=2\pi/P_{rot}$ is the 
angular velocity, and $R_A$ is the radius of the Alfven surface. 

For the early Sun at ages in the range $300-750\:$Myr, we adopt values 
for the rotation period of 5~days, the plasma density, where the outflow 
starts, is $10-20$ times solar, and a large-scale magnetic field 
strength $B=5\:$Gauss. In this case the quasi-steady mass loss rate is 
about $10^{-12}\:M_\odot$/year. This value can be increased by a factor 
of 2, if we take into account also the dynamic processes like coronal 
mass ejections which happened apparently more often at that era. Our 
estimates agree with those obtained by \citet{Cohen2014}.

\z{Mass loss} from the Sun is associated with quasi-steady plasma 
outflows from the corona. The hot coronal gas is concentrated in fairly 
low loops. The low-speed wind stream is formed near the top of the loop, 
in the cusp region, and it is enhanced slightly with an increase of the 
mass of the hot coronal plasma. Even if the soft X-ray radiation of the 
G-type star reaches the saturation level, i.e., $3-4$ orders of 
magnitude larger than the Sun, the rate of outflow of matter in the 
streamers does not increase by more than one order of magnitude. The 
high-speed flow from regions with open magnetic configuration is also 
increasing compared to the contemporary Sun. However, the high-speed 
wind of the young Sun must be amplified because the plasma density at 
the base coronal hole (or the polar region) increases, while the outflow 
is formed at higher coronal levels. It means that the contribution of 
the low-speed and high-speed streams to the quasi-steady mass loss can 
be similar. 

It is clear \z{that non-steady} processes were more frequent in the 
young Sun's corona.  Observations show that a large solar flare with the 
energy of about $10^{31}\:$erg is accompanied by CME ejected mass of 
about $10^{16}\:$g \citep{Drake2013}. We can estimate the CME mass loss 
by $\kappa^1\:$Cet if we turn to results of the \n{Kepler} mission, 
which discovered superflares on solar-mass stars \citep{Shibayama2013}. 
This study contains statistics on the occurrence frequency of the large 
flares. It shows that for $\kappa^1\:$Cet, the occurrence frequency of a 
flare with the energy of about $10^{34}\:$erg is 2 orders of magnitude 
higher than that for most active solar-type stars 
\citep[Fig.~9 in][]{Shibayama2013}. 
An extrapolation of this value to weaker 
phenomena with an energy of $10^{31}\:$erg under the same law as in 
Shibayama et al. gives us the occurrence frequency of such events around 
$(2-3)\times 10^{-26}\:$erg$^{-1}\:$star$^{-1}\:$year$^{-1}$. This leads 
to the value of $2\times 10^5$ events per year and corresponds to the 
CME mass loss of about $10^{-12}\:M_\odot$/year. Although this value is 
only 10\%\  of the possible rate of quasi-steady outflow of the young 
Sun, the contribution of CMEs could be significantly higher than today. 

Recently \citet{doNascimento2016} taking into account data on the 
large-scale magnetic field have been obtained the plasma outflow rate 
for $\kappa^1\:$Cet close to $10^{-12}\:M_\odot$/year. 
The similar value for this star was obtained 
by modelling of 3D-structure of stellar winds carried out by 
\citet{Fionnagain2019}. Improved detailed modelling of the CMEs 
emitted by $\kappa^1\:$Cet was made by \citet{Kay2019}.

Thus, the total mass loss of the young Sun was quite high, 
up to a \z{factor of} few times by 
$10^{-12}\:M_\odot$/year. 
If this rate is maintained for about a billion years, it \z{would} lead 
to a decrease in the mass of the Sun by 1\%. This does not affect the 
bolometric luminosity of the Sun, but maintains a high rate of decrease 
of the angular momentum.

\subsection{Remarks on stellar superflares}

The detection of extremely powerful non-steady phenomena, superflares, 
on \z{G-type} stars with the \n{Kepler} mission raises an alarming 
problem: can \z{super}flares happen on the present-day Sun, because 
their consequences can be catastrophic for our \z{contemporary high-tech 
civilization}. From the other hand, it gives a possibility to study 
flare activity of the Sun at different ages, i.e. at different rotation 
periods. 

The energy and occurrence frequency of \z{the largest flares 
(superflares)} on the young Sun can be evaluated from data provided by 
the \n{Kepler} mission \z{that had} monitored several hundred thousand 
stars during 10~years (2009 to 2018). In the short-cadence mode with the 
1-min temporal resolution 187 flares with the total energy from 
$2\times 10^{32}\:$erg to $8\times 10^{35}\:$erg were detected \z{from 
23~G-type} stars. One of the general conclusions of this mission was 
that only 0.2\%\  to 0.3\%\  of solar-type stars show superflares, while 
$>40\:$\%\  of the original solar-type superflare stars in previous 
studies are now classified as subgiants. The statistics of \n{Kepler'}s 
superflares \z{offered} the following estimates for the mean occurrence 
frequency of events: flares with the total energy $10^{33}\:$erg can 
occur once per $70-100\:$years, flares with the energy of $10^{34}\:$erg 
occur once in about $500-800\:$years, and superflares with the energy of 
$10^{35}\:$erg can happen once in about $4000-5000\:$years 
\citep{Maehara2015}. 

It is worthwhile to emphasize that stellar superflares with the total 
energies larger than $10^{35}-10^{36}\:$erg apparently occur on the 
youngest fast-rotating stars the saturated regime of activity, on 
subgiants and giants, \z{as well as} on components of close binary 
systems \citep{KaNi2018}.

An important question remains whether superflares with energies up to
$10^{35}-10^{36}\:$erg are possible \z{at all} on the Sun at present 
\z{time}. For comparison, the contemporary Sun demonstrates 1144 proton
flares \z{accompanied by proton events} with $E \ge 10\:$MeV during 
$1975-2003$ that corresponds to 41 events per year according to 
\z{an} IZMIRAN database \z{\protect\citep{Belov2005}.}

In order to evaluate the upper limit of the energy of flares \z{that} 
are able to occur in a given large active region, we carried out an 
analysis of observations of the total magnetic field vector, applying 
the method of the \z{non-linear force-free field} (NLFFF) extrapolation. 
\z{The NLFFF approximation gives information on the magnetic field structure 
in the corona of active regions (AR). It is assumed that the magnetic field 
above the photosphere can be regarded as a force-free field. Each AR has 
its specific magnetic configuration characterized by the energy of the 
magnetic field. The development of non-steady processes in ARs is 
governed by the difference of this magnetic energy from the energy of 
the potential field, but not by the total magnetic energy itself. This 
difference is characterized by the free energy. The amount of the free 
energy increases with the emergence of a new magnetic flux, and free 
energy is presumably responsible for flares and CMEs. If one considers 
only the total energy of non-steady processes, it is possible to shift 
from the equations describing the structural features to considering the 
magnetic virial theorem. An analytic expression for the free energy of 
the solar corona as a whole was suggested earlier by \citet{LiRuKaMy2015}
which let us assess an absolute upper estimate of the energy of 
flares that are possible for a given large active region.} We found that 
even the largest active regions on the Sun are not able to produce 
non-steady processes like flares and CMEs with the total energy greater than 
$3\times 10^{32}\:$erg. \z{Consequently,} flares stronger than this value 
cannot occur on the contemporary Sun 
\protect\citep{LiRuKaMy2015,KaLi2015}. \z{This is approximately equal to 
the total energy estimated for the Carrington event, 
$2\times 10^{32}\:$erg.}

Considering the star $\kappa^1$~Cet as an analog of the young Sun, 
\z{one could} derive the following parameters of its activity: the X-ray 
luminosity is $L_X =10^{29}\:$erg/s, occurrence frequency of flares with 
the total energy $E > 10^{32}\:$erg is 5~events per day or 1825~events 
per year (from \n{EUVE} data by \citealt{Audard2000}), \z{and} its surface 
is more spotted (by a factor of $10-20$). The average longitudinal 
magnetic fields of G stars (at an age of $0.6-1\:$~Gyr) are $10-15$ times 
stronger than the maximum magnetic field of the Sun as a star (as it 
follows from the results by \n{the Bcool collaboration}, see the 
previous section). Therefore, because the total flare energy that is 
\z{proportional} to $B^2$, the maximal possible flare energy of young 
G-\z{type} stars cannot exceed $5\times 10^{34}\:$erg. 
It is consistent with the statistics for the \n{Kepler} 
targets \z{among the main-sequence stars} with the rotation periods
around 10 days. Only flares with these total energies can be considered
as solar-type analogs of impulsive events, associated with the energy
deposit in and above the chromosphere \z{with} its subsequent release, 
as it happens on the Sun. \z{So}, the stronger events require for their 
explanation, either non-solar analogies for the origin of flares or 
appropriate changes in the dynamo mechanism \citep{KaEtal2018}. 
\z{From the other hand, taking into account new results by 
\protect\citet{Kochukhov2020} on real unsigned small-scale strong 
magnetic fluxes with a big filling factor on a surface of young 
solar-type stars, it could be possible to obtain also 
more energy for superflares. 
}

\section{Discussion}

\z{In this study we summarize the key points} of the scenario of the
evolution of activity: 

\begin{enumerate}[\textbullet]
\item the evolution of solar-stellar activity follows 
the temporal behavior of the angular momentum of a star. 
\z{A joint consideration of the rotation-age and activity-rotation 
relationships gives a possibility to explore of basic indicators of 
activity for solar-type stars of various ages.}

\item \z{As demonstrated in the} empirical trends traced in recent studies 
of stellar magnetism, after the $100-250\:$Myr, magnetic fields evolve in 
a close correlation with rotation and determine the pattern of 
activity over the next billions of years. 

\item It is now clear that the lifetime of the Sun on the main sequence can be 
divided \z{in} a few epochs: the early Sun when the Solar Planetary System 
just was forming, when the solar rotation rate was $10-20$ times faster than 
nowadays; the era of the young Sun when a regular cycle was established 
(the rotation rate was $2-5$ times faster), and the contemporary epoch 
of the slow-rotating Sun. \z{Each of these epochs is characterized by 
different behavior of the basic tracers of activity.}

\item The activity of the early Sun is \z{similar} to young solar analogues, 
showing a very high level at the saturated regime. It is characterized 
by large active region area, a dense, active coronae with the maximum 
level of the coronal activity $\left(\log L_X/L_{bol}=-3\right)$, 
superflares, enhanced lithium abundance, \z{and} strong plasma outflows. 
Saturation is also present in large-scale magnetic fields. This epoch is 
characterized by independence of the main activity tracers on rotation.

\item As \z{the rotation} of solar-mass main sequence stars \z{is slowed} by 
its magnetized wind, the saturated regime of activity changes to the 
unsaturated mode, when activity indices decline in tandem with rotation. 
This solar-type activity with formation of the regular \z{activity} 
cycle, as oscillating dynamo does turn on, occurs when the rotation 
period of a G-\z{type} star (the young Sun) reaches a value of about few 
days.
\end{enumerate}

\section{Conclusions}

This \z{study briefly summarizes the} results of many laborious long-term
observational programs \z{conducted} by large scientific teams \z{and}
dedicated to investigations of different aspects of stellar activity. 
\z{Their results are considered} in the context of general ideas, which
\z{facilitate an} understand\z{ing} the evolution of solar activity from
its \z{arrival} on the main sequence through the epoch of the sunspot
cycle formation up to the present days. Basic astrophysical laws,
derived for the rotation-age and activity- rotation relationships,
allow\z{ed} us to trace how activity of low-mass stars changes with age
during their stay on the main sequence. \z{Recent results} clarify,
complement and expand the general ideas on the evolution of solar and
stellar activity. We focus on the activity properties of stars that can
act as proxies for the early and the young Sun. Our analysis includes
joint consideration of different tracers of activity, rotation and
magnetic fields of Sun-like stars of various ages and helps to
understand the physics of these phenomena.

We \z{identify} rotation periods, when the saturated regime of activity
changes to the unsaturated mode, \z{when} the solar-type activity 
\z{is formed}: for G- and K-\z{type} stars, they are 1.1 and 3.3 days,
respectively. This corresponds to \z{an} age interval of about 
$0.2 - 0.6\:$Gyr, when \z{regular} sunspot cycle beg\z{a}n to be 
established on the early Sun. 

Properties of the coronal and chromospheric activity of the early and 
the young Suns are discussed. Our analysis of the EUV-fluxes in the 
spectral range of $1350 - 1750\:$\AA\  showed that the far-UV radiation 
of the early Sun was by a factor of 7~more intense than that \z{of} the 
present-day Sun, and twice higher than when the regular sunspot cycle 
was established. 

For the young Sun, we can estimate the possible mass loss rate
associated with quasi-steady outflow as $10^{-12}\:M_\odot$/yr. Studies
of the largest flares on solar-type stars have \z{also been discussed
and we concluded} that the most powerful phenomena occur on 
fast-rotating stars in the saturated activity regime. Our estimate 
of stellar magnetic fields \z{offers} a possibility \z{to estimate} the
maximum possible flare energy. This value for flares on young 
main-sequence G-type stars with the rotation period 
$P_{rot}\sim 8-10\:$d and ages 
600~Myr--1~Gyr was \z{assessed as not be in excess} of 
$5\times 10^{34}\:$erg. For such events, we conclude that their 
origin is similar to solar one: \z{when the free energy is accumulated
in the chromosphere and then this deposited energy is releazed} 
in the course of a non-steady process. 

As for superflares on the Early Sun with the total energy 
$10^{35}-10^{36}\:$erg, \z{the} mechanism of \z{such} phenomena
obviously differs from that accepted for contemporary solar flares, 
and \z{possibly} can be associated with the evolution of \z{the} 
large-scale magnetic fields or another dynamo regime. 
Further study of these problems \z{should} help \z{in} understand\z{ing}
the origin of extreme events on the Sun in the past, \z{which is
important for establishing the physical} conditions in the heliosphere
from an epoch of formation of the solar planetary system, in general,
the early Earth and, in particular, the origin of terrestrial biosphere.

\section{Acknowledgements}

I am grateful the SOC of VarSITI Completion General Symposium 
(held in Sofia, Bulgaria on June 10-14, 2019) for \z{the invitation and}
possibility to present these results. 
I am thankful to anonymous reviewers for helpful remarks 
that help to improve this paper.

I am deeply grateful to Dr. S. Saar for useful discussions and comments. 

This work was carried out with the partial support by the 
Russian Foundation for Basic Research (project numbers 19-02-00191a 
and 18-52-06002 Az\_a).
This research has also made use of NASA's Astrophysics Data System.

\bibliographystyle{cas-model2-names}

\bibliography{jastp-mk}

\end{document}